\documentclass{elsart}
\usepackage{epsfig}
\usepackage{graphicx}
\usepackage{hyperref}
\usepackage{bbold}

\journal{Nuclear Instruments and Methods A}

\begin{document}

{\tt PITHA 04/01\\}
\vspace*{1cm}

\begin{frontmatter}
\title {Charge Transfer and Charge Broadening\\
        of GEM Structures in High Magnetic Fields}

\author[A]{M.~Killenberg},
\author[A]{S.~Lotze},
\author[A]{J.~Mnich},
\author[A]{A.~M\"unnich},
\author[A]{S.~Roth\thanksref{cor}},
\author[B]{F.~Sefkow},
\author[A]{M.~Tonutti},
\author[A]{M.~Weber},
\author[B]{P.~Wienemann}
\address[A]{Rheinisch-Westf\"alische Technische Hochschule, D-52056 Aachen, Germany}
\address[B]{Deutsches Elektronen-Synchrotron, D-22607 Hamburg, Germany}

\thanks[cor]{Corresponding author.
              {\it Phone:} +49--241--80--27296; 
              {\it Fax:} +49--241--80--22244;
              {\it Email:} {\tt roth@physik.rwth-aachen.de}}

\begin{abstract}
We report on measurements of charge transfer in GEM structures
in high magnetic fields.
These were performed in the framework of the R\&D work for a 
Time Projection Chamber at a future Linear Collider.
A small test chamber has been installed into the aperture of a 
superconducting magnet with the GEM structures mounted
perpendicular to the B field direction.
The charge transfer is derived from the electrical currents 
monitored during irradiation with an ${}^{55}$Fe source.
No severe loss of primary ionisation charge is observed,
but an improved ion feedback suppression is achieved for 
high magnetic fields.
Additionally, the width of the charge cloud released by
individual ${}^{55}$Fe photons is measured using a finely
segmented strip readout after the triple GEM structure.
Charge widths between 0.3 and 0.5~mm RMS are observed,
which originate from the charge broadening inside the
GEM readout. 
This charge broadening is only partly suppressed at 
high magnetic fields.
\end{abstract}


\end{frontmatter}

\section{Introduction}
\label{introduction}

A Time Projection Chamber (TPC) is foreseen as the main tracker
of the detector for a Linear Collider such as TESLA~\cite{tdr}.
Currently the possibility to use Gas Electron Multipliers (GEM) \cite{sauli} 
for the charge amplifying system is studied extensively.
When using GEMs for the TPC end plate, the pads directly detect the amplified
electron cloud which results in a fast and narrow charge signal. 
The fine structures of the GEM of the order of 100~$\mu$m size allow
the detection of the charge cloud with high spatial resolution.
In contrast to wires, a GEM shows no preferred direction, thus any 
$\vec{E} \times \vec{B}$ effects will be isotropic. 
Finally, using different electric fields on both GEM sides, 
the drift of ions produced inside the GEM holes into the TPC drift volume 
(ion feedback) can be suppressed and the need of a gating grid can be 
potentially avoided.

To demonstrate the advantages of a TPC with GEM readout,
a prototype chamber will be built within the R\&D activities of 
the Linear Collider TPC group~\cite{tpc-prc,lctpcrd} resembling
a sector of the TPC as in the TDR design.
It will be operated in test beams and in magnetic fields up to 5~T.
The end plates of this prototype will contain up to three planes
of charge amplifying GEM foils.
The operation conditions of this multi-GEM-structure have to be
studied beforehand, because in the case of three GEM planes six
electric fields have to be set.
For these studies two small test chambers were set up at RWTH Aachen
and installed in a 5~T magnet at DESY Hamburg.
The first chamber was dedicated to the measurement of charge transfer
in high magnetic field and the second one to the determination of the
transverse charge spread as a function of the magnetic field.

\begin{figure}[b]
\begin{center}
  \includegraphics[height=0.3\textheight]{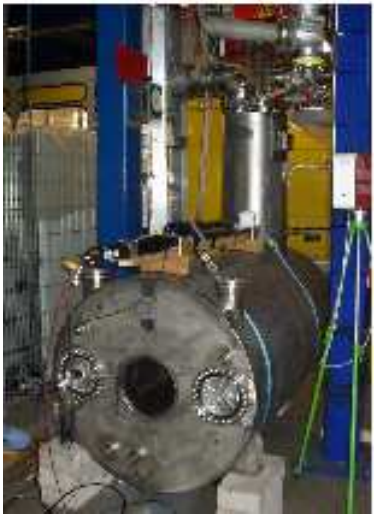}
  \hfill
  \includegraphics[height=0.3\textheight]{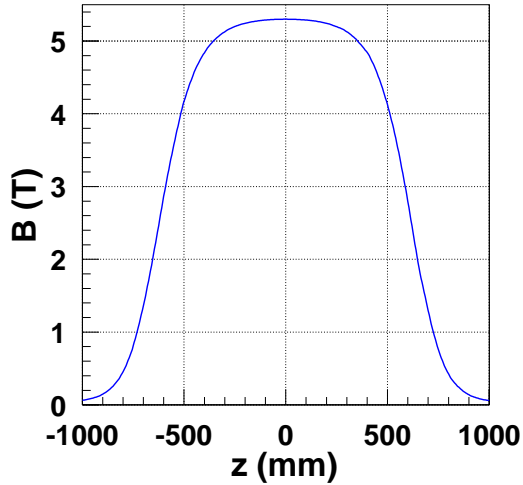}
\end{center}
\caption{The superconducting 5~T magnet in the HERA cryogenic plant
and its field strength along the magnetic axis at a coil current of 1000~A.}
\label{fig:magnet}
\end{figure}

\section{Experimental Setup}
\label{setup}

The measurements described below were accomplished in a 5~T 
superconducting magnet at DESY (see Figure~\ref{fig:magnet}). 
After formerly serving as magnet compensating the
main field of the ZEUS experiment, the solenoid has
been reinstalled in the cryogenic plant of the HERA accelerator
at DESY to provide a high magnetic field facility for
detector R\&D projects for a future Linear Collider such as TESLA.

The magnet has an aperture of 28~cm and a coil length of 120~cm.
Including its cryostat, the total length amounts to 186~cm. 
The coil is cooled by 4.5~K liquid helium to ensure superconductivity.
The solenoid is operated at currents up to 1000~A corresponding 
to a central magnetic field of up to 5.3~T as shown in Figure~\ref{fig:magnet}. 
With the ramp rate of 0.5~${\rm As}^{-1}$, it takes about half 
an hour to ramp up the magnet from 0~T to 5.3~T. 
More detailed information on the magnet facility is presented in \cite{magnet}.

\begin{figure}[b]
 \begin{center}
  \includegraphics[width=0.49\textwidth]{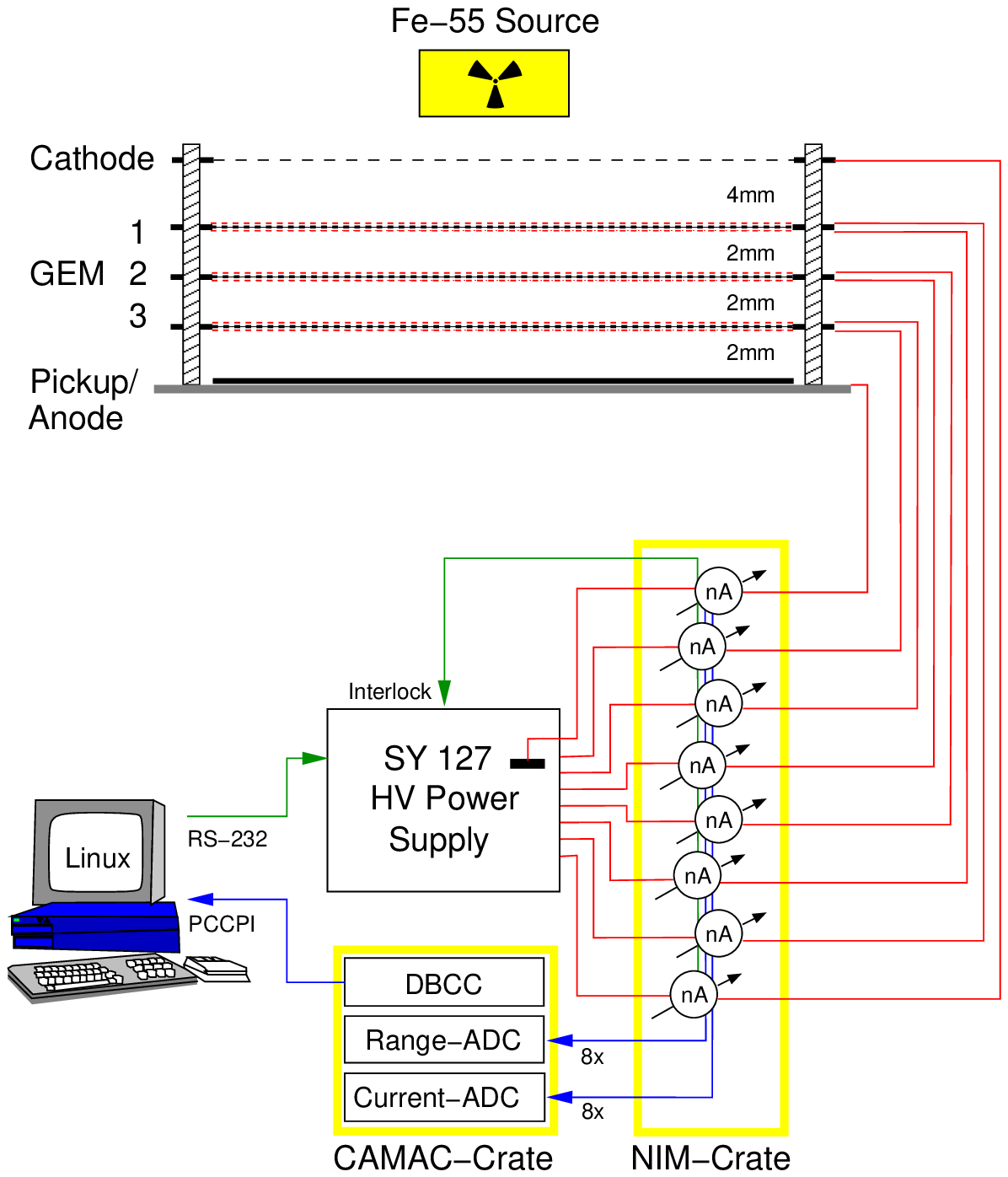}
  \hfill
  \includegraphics[width=0.49\textwidth]{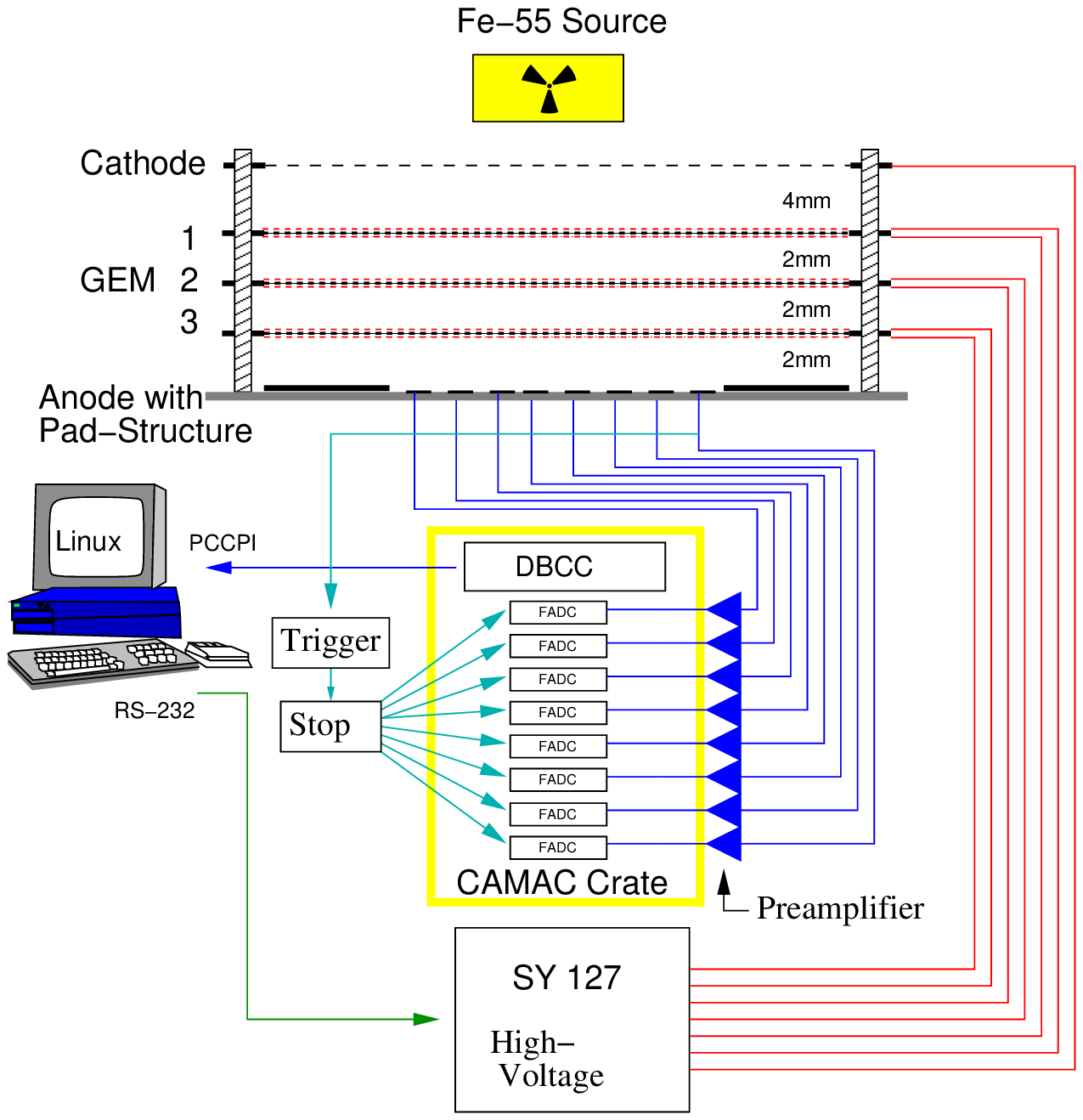}
  \caption{Test chamber setup for measurements of the 
           charge transfer (left) and of the charge spread (right)}
  \label{fig:setup}
 \end{center}
\end{figure}

The mechanical and electric setup of our test chambers has been
described in detail in Reference~\cite{ourpaper}.
The gas volume consists of a composite frame enclosing a
stack of three standard $10 \times 10~{\rm cm}^2$ GEM foils.
Using thin absorbers the radiation from an ${}^{55}$Fe source of 1~GBq
activity is diminished such that about $2 \cdot 10^6$ photons per 
second reach the chamber.
A small window on top of the chamber covered by a 24~$\mu$m thick
mylar foil allows the photons to penetrate into the chamber.

The chambers were operated with a gas mixture as it is proposed 
in the TESLA technical design report~\cite{tdr} which consists
of Ar(93\%),CH${}_4$(5\%),CO${}_2$(2\%).
A regulation system was used to keep the absolute pressure of the 
chamber at a constant value within 1~mbar.
In addition the chamber's temperature was continuously monitored
and recorded.
Each electrode (GEM surfaces and cathode) is connected to an 
individual channel of a CAEN SY127 HV power supply.

For the current measurements the anode plane consists of a solid 
copper electrode of the same size as the GEM structures \cite{sven}.
Current monitors with a resolution of about 0.1~nA 
are inserted into the supply line of each HV channel. 
To measure the anode current, the anode plane is connected to 
ground via an additional current monitor. 
The high voltage control and current readout are handled by a 
custom application running on a Linux PC.

For the measurement of individual pulses an anode with a finely 
segmented area with 8 strips of 0.8~mm pitch is used \cite{astrid}.
The signal pulse of each strip is read out via a preamplifier
and digitised using a 100~MHz Flash ADC.
The eight FADCs digitise their input signal continuously and store
it into a ring-buffer memory.
After arrival of a trigger all FADCs are stopped and the
memory is read out by a Linux computer.
A trigger is generated if one of the strip signals exceeds the
level of an analog discriminator.

\begin{figure}[b]
 \begin{center}
  \includegraphics[width=0.8\textwidth]{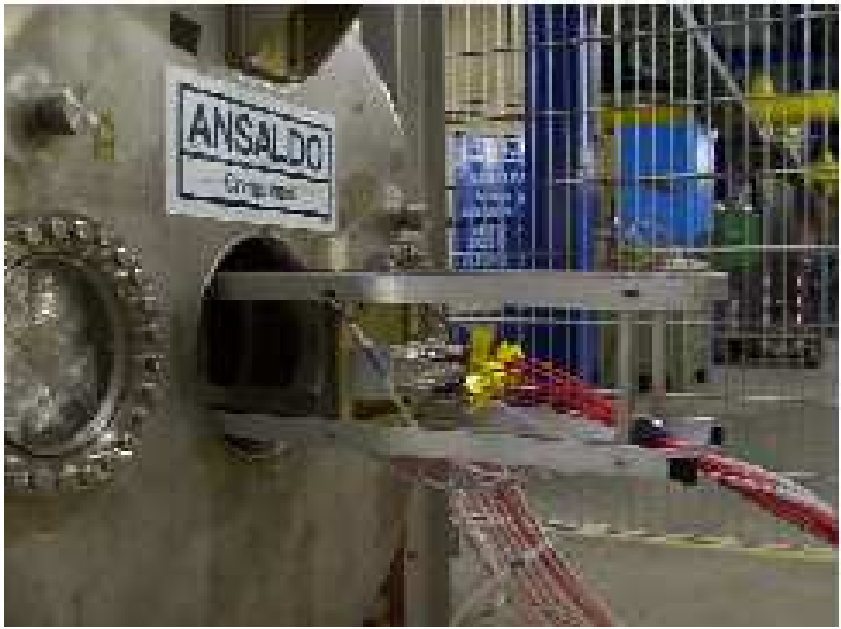}
  \caption{Insertion of the test chamber into magnet}
  \label{fig:chamber}
 \end{center}
\end{figure}

In Figure~\ref{fig:setup} an overview of the electric setup 
and read-out system of the two test chambers is given.
Figure~\ref{fig:chamber} shows one of the chambers just before
insertion into the aperture of the superconducting magnet.

\clearpage

\begin{figure}[!ht]
 \begin{center}
  \includegraphics[width=0.45\textwidth]{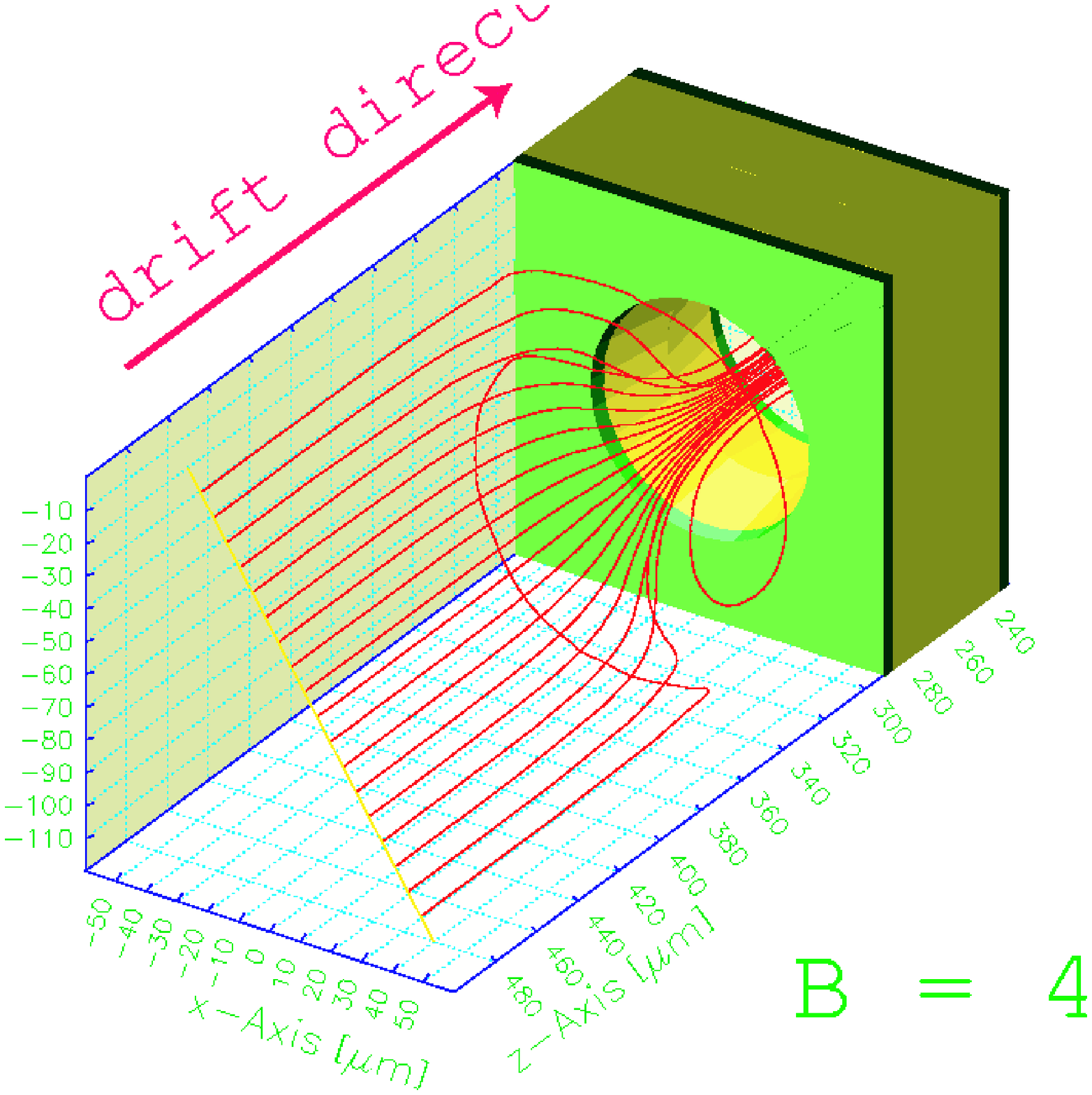}
  \caption{GARFIELD simulation of electron drift lines starting in front of a GEM;
           at 4~T magnetic field no electrons are lost during collection 
           into the GEM holes.}
  \label{fig:collection}
 \end{center}
 \begin{center}
  \vspace{0.5cm}
  \includegraphics[width=0.45\textwidth]{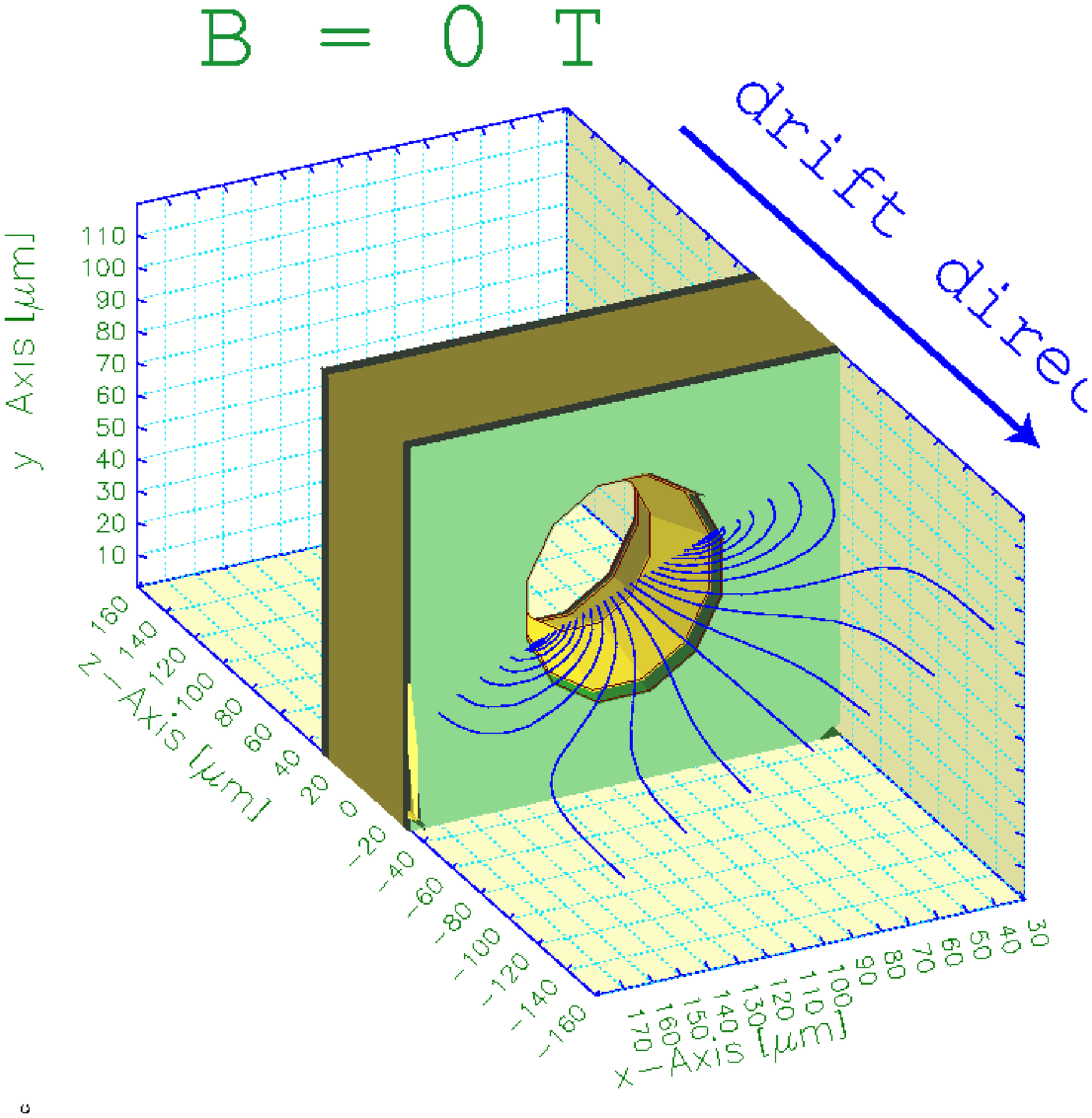}
  \hfill
  \includegraphics[width=0.45\textwidth]{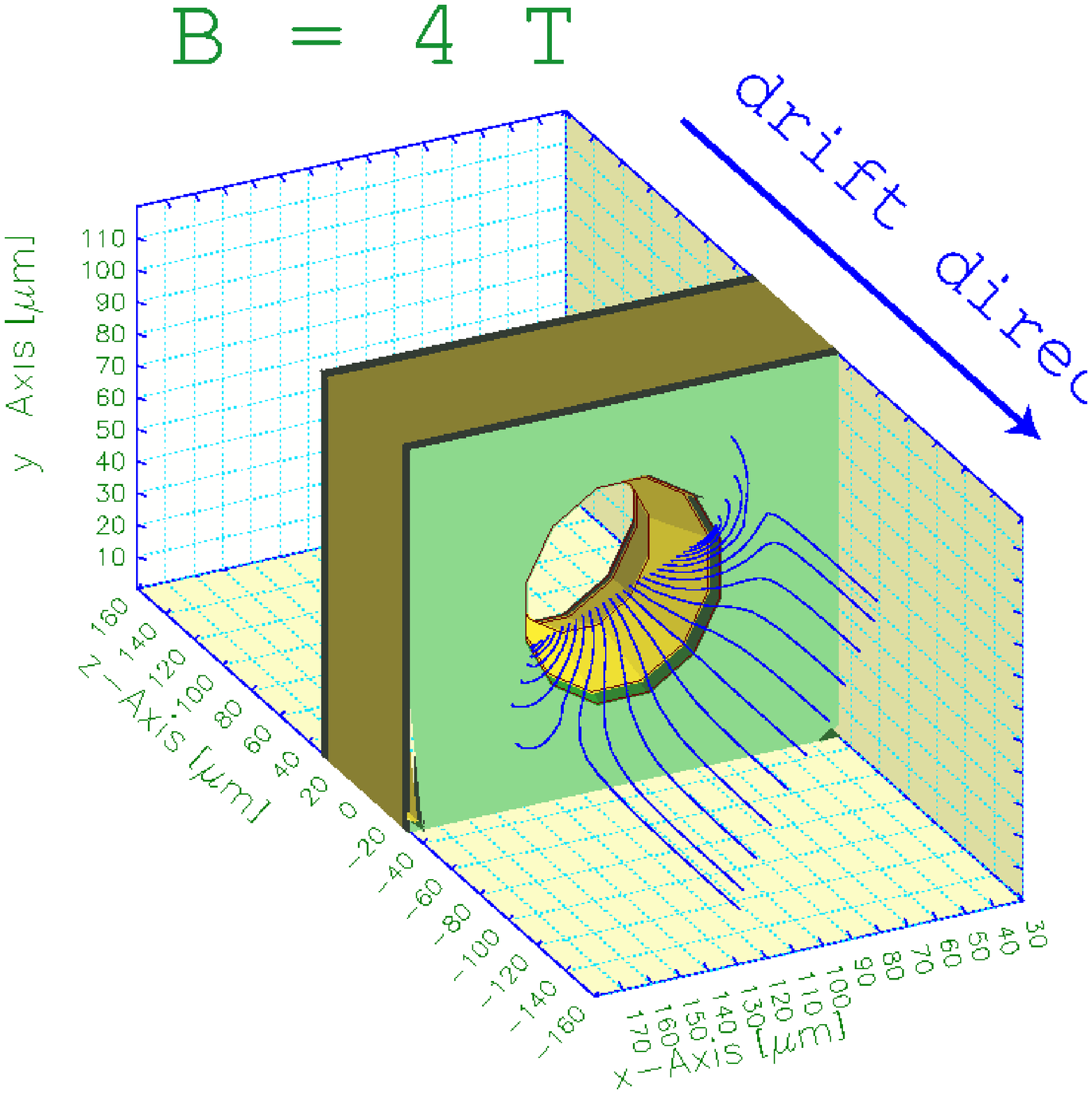}
  \caption{GARFIELD simulation of electron drift lines starting inside a GEM hole;
           extraction is improved in 4~T field
           (right) as compared to no magnetic field (left).}
  \label{fig:extraction}
 \end{center}
\end{figure}

\section{Measurement of Charge Transfer in High Magnetic Fields}
\label{transfer}

Our primary motivation for making tests in a magnet was to investigate
whether there is a significant drop in collection efficiency for GEMs
in high B-fields parallel to the electric field as suggested by the
Langevin formula
\begin{equation}
\label{langevin}
\vec{v}_{Drift} \propto \hat{\vec{E}}
     +\omega\tau\;(\hat{\vec{E}}\times\hat{\vec{B}})
     +\omega^2\tau^2\;(\hat{\vec{E}}\cdot\hat{\vec{B}})\;\hat{\vec{B}}~,
\end{equation}
where $\hat{\vec{E}}$ and $\hat{\vec{B}}$ are unit vectors of the fields,
$\tau$ is the mean time between two collisions of the drifting electron,
and $\omega = eB/m$ is the cyclotron frequency.
The last term proportional to $\omega^2$, which gives the contribution
along the magnetic field lines, could cause a drop in collection
efficiency for high magnetic fields. When this term dominates, most
electrons will no longer be collected into a GEM hole, but stay
on drift lines perpendicular to the GEM surface and eventually reach
the GEM's copper coating.
Those charges would be lost for the signal and consequently decrease
the chamber's ${\rm d}E/{\rm d}x$ capabilities due to the loss in primary
ionisation statistics.

Figure~\ref{fig:collection} shows a numerical simulation using the 
programs MAXWELL~\cite{maxwell}, MAGBOLTZ~\cite{magboltz} and
GARFIELD~\cite{garfield}.
In total 20 drift lines of electrons starting in front of a GEM hole 
are shown for a 4~T magnetic field perpendicular to the GEM surface. 
None of the drift lines end up on top of the GEM, but all electrons
are collected into the GEM hole along curling tracks.
From these simulations no significant drop in collection efficiency 
is expected in high magnetic fields.
On the other hand the numerical simulations predict an increase in extraction 
efficiency.
The situation is shown in Figure~\ref{fig:extraction}, where the
drift lines of electrons out of a GEM hole at 0~T and at 4~T magnetic field
are compared.
In both simulations the same number of electrons start within the GEM hole.
From the number of extracted drift lines one observes that more of the electrons 
are extracted at 4~T.
This is explained by the third term in Equation~\ref{langevin} which is 
bending the drift lines along the B field direction.

\begin{figure}[!hp]
 \begin{center}
  \includegraphics[width=\textwidth]{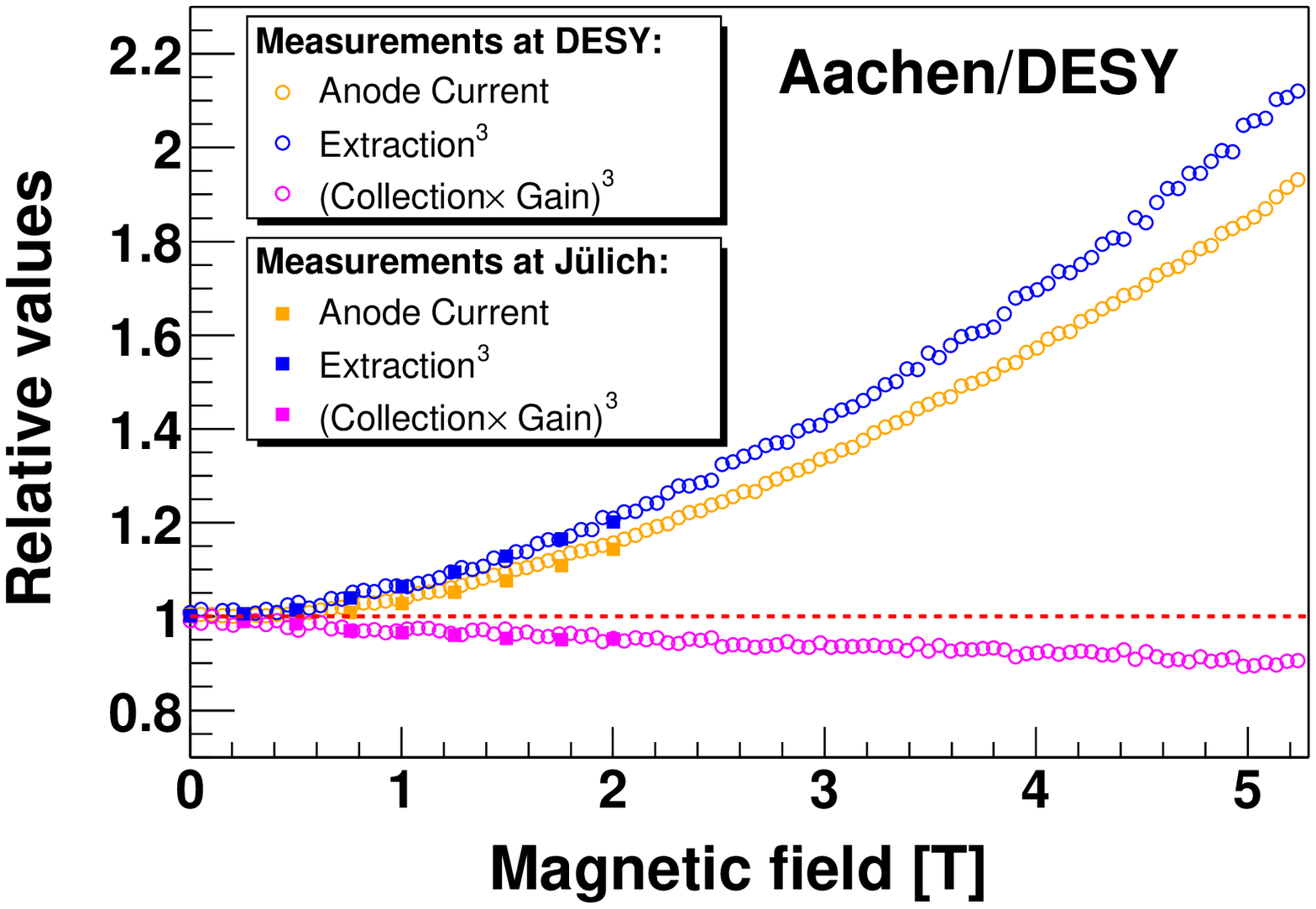}
  \caption{Measurement of anode current and extraction efficiency
           versus the magnetic field; From these curves the product 
           of collection and gain is calculated.}
  \label{fig:juelich}
 \end{center}
 \vspace*{1cm}
 \begin{center}
  \includegraphics[width=\textwidth]{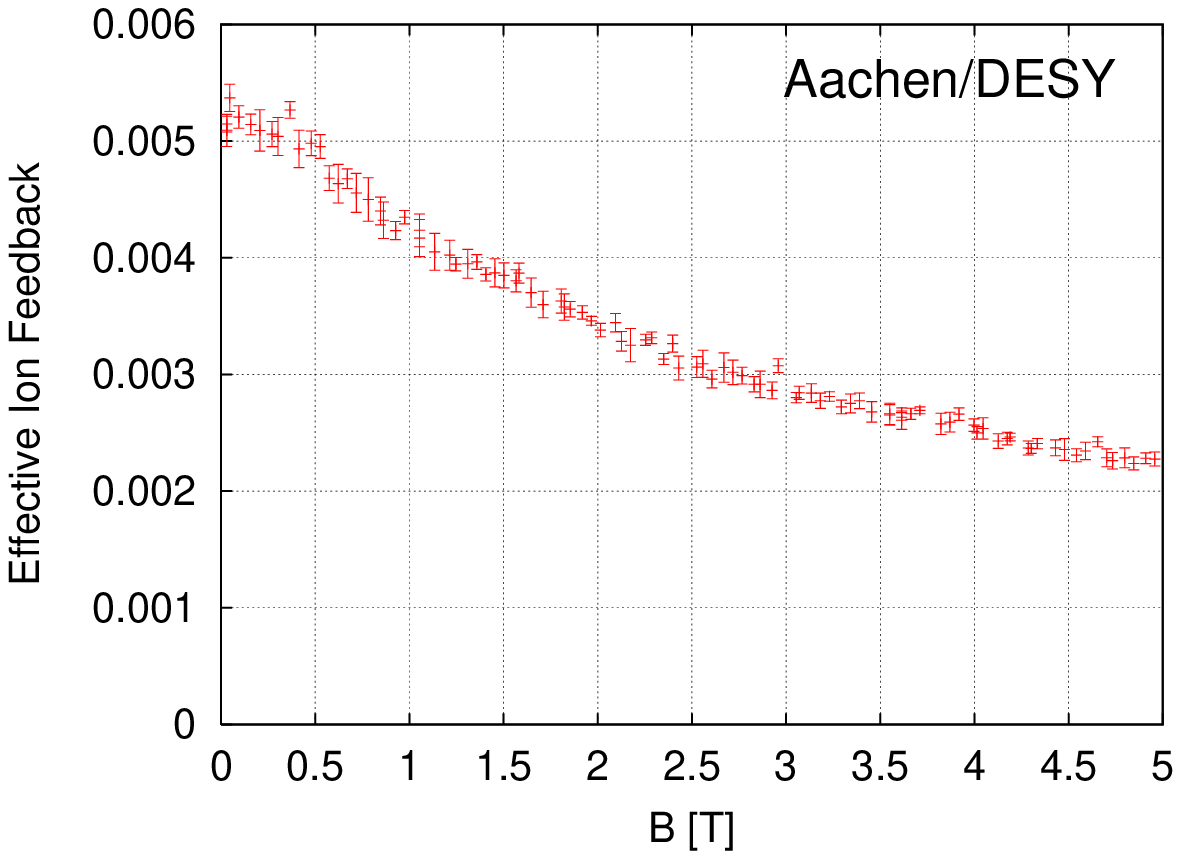}
  \caption{Measured ion feedback with optimised chamber settings
           vs.\ magnetic field}
  \label{fig:ionfeedback}
 \end{center}
\end{figure}

In Figure~\ref{fig:juelich} the measured anode current as
function of the magnetic field is shown.
Measurements done in a 2~T magnet at the Forschungszentrum J\"ulich 
\cite{ourpaper} are nicely reproduced by our new measurements
at the 5~T magnet at DESY.
It rises approximately by a factor of 2 between 0~T and 5~T.
Simultanously the extraction efficiency of the electrons
out of the last GEM in front of the anode plane was measured.
Because the chamber was operated with a symmetric setup 
(all GEM voltages and electric fields at the same value)
the increase of the anode current can be calculated from
the product of collection efficiency, gain and extraction
efficiency of a single GEM to the power of three.
The independently measured extraction efficiency 
can explain the rise in the anode current.
This shows that the product of collection efficiency and gain
is only slightly affected by the change of the magnetic field.
We see no indication for severe losses of primary ionisation charge.

One important charge transfer quantity is the ion feedback, 
which describes how much ion charge is transferred into the 
drift volume per electron charge collected on the anode plane.
Ions reaching the TPC drift volume would represent
a positive space charge and deteriorate the electric drift field. 
The ion feedback can be derived from the measurements by
dividing the cathode current by the anode current.
Highly asymmetric electric fields on both sides of a 
GEM lead to ion suppression.
The small electric field in the drift volume for example
causes many of the drift lines from the amplification 
region inside the GEM hole to end on the copper plane facing the
TPC drift volume.

Therefore, the relatively small drift field typical of a TPC
automatically leads to ion feedback suppression.
Moreover, the ion feedback can be minimised by the variation of 
the electrical fields within the GEM structure~\cite{nim-if}.
We searched for a setting with optimum ion feedback suppression
with the drift field constrained at 200~V/cm
and requiring an effective gain of $10^4$.
The first two GEMs are set at voltages of 310~V, the last GEM
at 350~V.
The first transfer field and induction field are set very high
(6000~V/cm and 8000~V/cm respectively) whereas the second transfer 
field is set rather low (60~V/cm).
Using this optimised setting allows a long term and stable operation of 
a triple-GEM structure with ion feedback well below 1\%.
The ion feedback is then measured as a function of the magnetic field.
The result is presented in Figure~\ref{fig:ionfeedback} and shows
that the ion feedback is further reduced with increasing magnetic field strength.
This is due to the improved electron extraction already mentioned before.
In a magnetic field of 4~T an ion feedback of only 2.5 permille
has been achieved.

\section{Dependence of Charge Width on the Magnetic Field}
\label{width}

The information on the charge spread within a GEM structure
is important to estimate the optimum pad size for the GEM TPC.
The spatial resolution of a TPC improves with a magnetic field
parallel to the electric field, because the transverse diffusion of
the electron cloud is suppressed.
A very narrow primary charge cloud, for example 300~$\mu$m in
Ar(95\%):CF${}_4$(5\%) at 4~T after a drift distance of 2.5~m, 
will reach the GEM readout structure.
Then the charge cloud is further broadened by effects
within and between the GEM foils.
Possible effects are the gas amplification, highly divergent 
electrical field lines and the diffusion within high electric fields.
The part of the charge broadening which 
is caused by diffusion effects can be reduced in high magnetic fields.
We present new measurements of the charge spread by GEM structures 
in high magnetic fields.

The cluster width distribution for individual events has 
been analysed using a segmented pad anode.
Only events of the photo peak, where the total photon
energy has been deposited inside the chamber, are used
for this analysis.
In these events the 6~keV photon of the ${}^{55}$Fe source 
is converted into a 3~keV photo-electron, which is predominantly
emitted perpendicular to the incoming photon.
Additionally, a 3~keV Auger electron is emitted from the
excited argon ion.
The two electrons ionise the gas along their tracks
and a primary charge cloud is produced.
The primary charge cloud reaches the GEM structure after
a maximum drift distance of 4~mm.
Then the width of the charge cloud is further broadened 
inside the GEM structure.

\begin{figure}[!hp]
 \begin{center}
  \includegraphics[width=\textwidth]{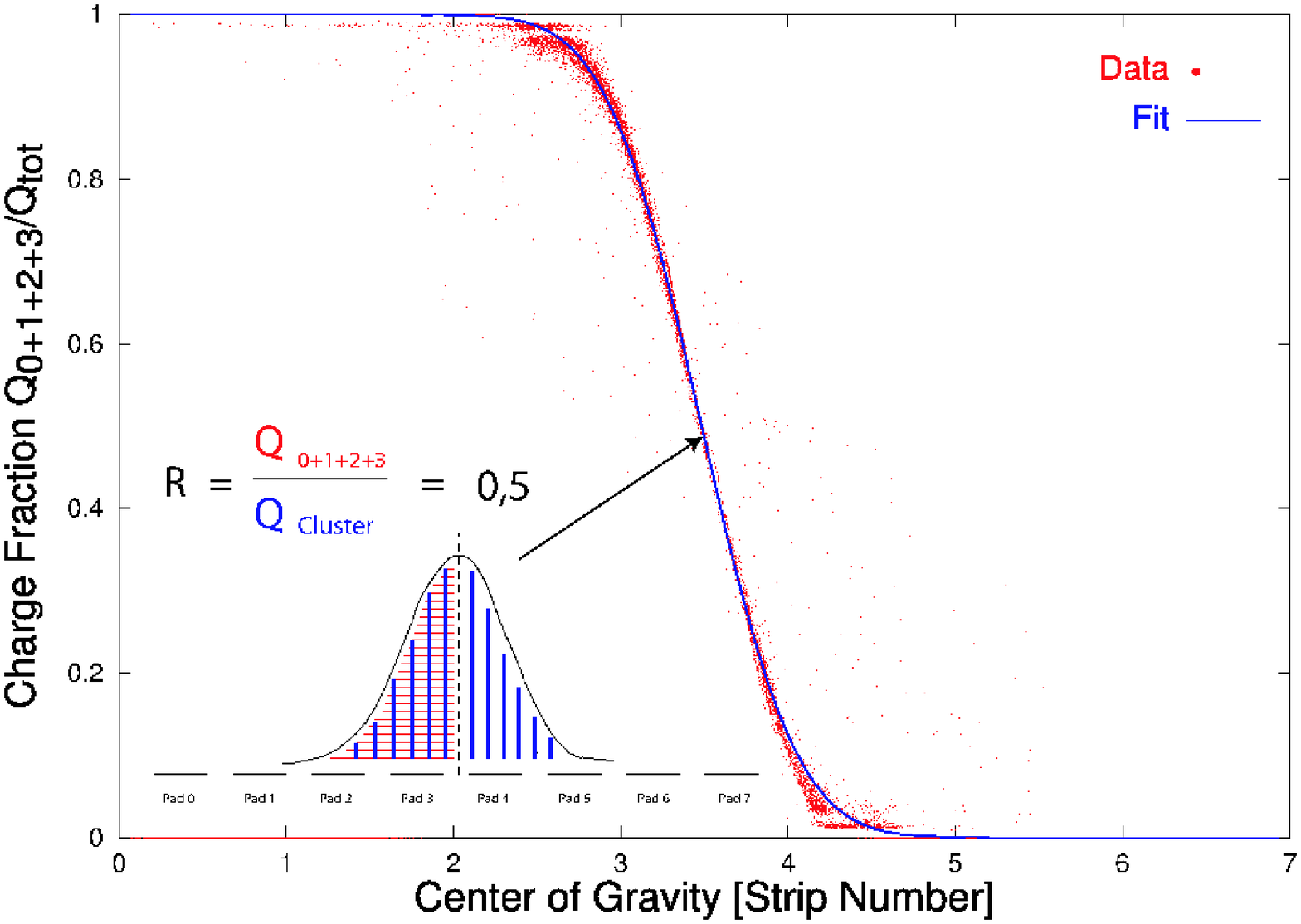}
  \caption{Charge distribution on strips}
  \label{fig:eta}
 \end{center}
 \vspace*{1cm}
 \begin{center}
  \includegraphics[height=\textwidth,angle=-90]{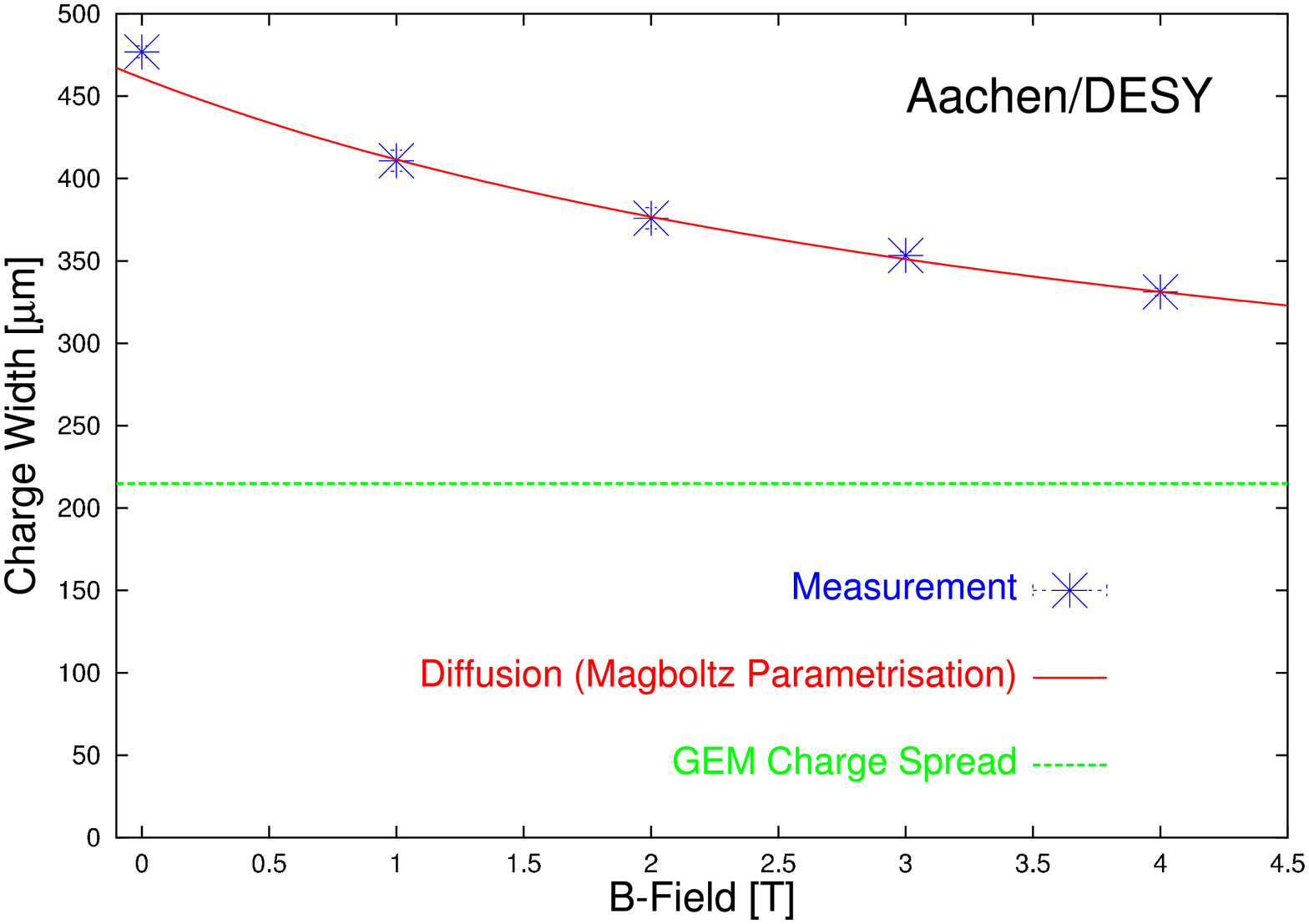}
  \caption{Cluster width vs.\ the magnetic field; the measurements
           are compared with a quadratic sum of 
           a B-field independent charge spread and the diffusion as
           predicted by the MAGBOLTZ program.}
  \label{fig:spread}
 \end{center}
\end{figure}

In this experiment the width of the charge cloud
is determined from the charge sharing between adjacent strips.
First the centre of the charge cloud is calculated from
the centre of gravity of the measured signals on each of 
the eight strips.
The charge sum measured on the left half of the anode
(strips 0--3) is compared to the charge sum on
the right side (strips 4--7).
The charge ratio, 
\begin{equation}
   R = \frac{Q_{\rm left}}
            {Q_{\rm left} + Q_{\rm right}} \; ,
\end{equation}
is calculated for each event.
This ratio is then plotted versus the charge centre
in Figure~\ref{fig:eta}.
As result one observes the typical shape of an error function
as expected for a charge cloud of gaussian shape.
From the slope of the data points one determines the width 
of the charge distribution.
The charge width of the primary ionisation cloud is estimated
to be in the range of 30--70~$\mu$m, depending on the B field, 
and can therefore be neglected.

The result of the cluster width measurement is shown in Figure~\ref{fig:spread}. 
The RMS width of the charge cloud is reduced from 0.5~mm without magnetic field
to 0.3~mm at 4~T.
From a numerical simulation using the program MAGBOLTZ we would expect 
a charge width between 400~$\mu$m at 0~T and 240~$\mu$m at 4~T due to
the diffusion in the high transfer fields between the GEM foils.
If we add a B-field independent charge broadening effect
of the GEM structure of 215~$\mu$m size quadratically, 
measurement and simulation agree.

\section{Conclusions}
\label{conclusions}

The charge transfer and the charge broadening of triple-GEM structures
have been studied in magnetic fields up to 5~T perpendicular to the
GEM surface.
With increasing magnetic field, an improved electron extraction efficiency
has been measured, leading to a reduction of ion feedback by a factor
of 2.
No indication for a large drop in electron collection 
efficiency has been observed.
The charge broadening of GEM structures has been measured and we
conclude that the charge broadening nature of a triple GEM
structure is not severely diminished by magnetic fields up to 4~T.

\section{Acknowledgements}
\label{acknowledgements}

We thank the CERN printed-circuits workshop and R.~Oliveira for
providing the GEM foils.
Those of us from RWTH Aachen thank DESY for its hospitality.
We acknowledge the financial support of this R\&D project by the
German Bundesministerium f\"ur Bildung und Forschung.

\end{document}